\documentclass[sigconf,natbib=true]{acmart}
\usepackage[ruled,linesnumbered]{algorithm2e}
\usepackage{multirow}
\usepackage{pifont}

\renewcommand\footnotetextcopyrightpermission[1]{} 
\AtBeginDocument{%
  }

\setcopyright{acmcopyright}
\copyrightyear{2024}
\acmYear{2024}
\acmDOI{XXXXXXX.XXXXXXX}

\begin{document}
\begin{sloppypar}
\fancyhead{}

\title{MealRec$^+$: A Meal Recommendation Dataset with Meal-Course Affiliation for Personalization and Healthiness}

\author{Ming Li}
\email{liming7677@whut.edu.cn}
\affiliation{
  \institution{Wuhan University of Technology}
  \city{Wuhan}
  \country{China}
}

\author{Lin Li}
\email{cathylilin@whut.edu.cn}
\authornote{Corresponding author}
\affiliation{
  \institution{Wuhan University of Technology}
  \city{Wuhan}
  \country{China}
}

\author{Xiaohui Tao}
\email{Xiaohui.Tao@unisq.edu.au}
\affiliation{
  \institution{University of Southern Queensland}
  \city{Toowoomba}
  \country{Australia}
}

\author{Jimmy Xiangji Huang}
\email{jhuang@yorku.ca}
\affiliation{
  \institution{York University}
  \city{Toronto}
  \country{Canada}
}

\renewcommand{\shortauthors}{Ming Li et al.}

\begin{abstract}

Meal recommendation, as a typical health-related recommendation task, contains complex relationships between users, courses, and meals. 
Among them, meal-course affiliation associates user-meal and user-course interactions. 
However, an extensive literature review demonstrates that there is a lack of publicly available meal recommendation datasets including meal-course affiliation. 
Meal recommendation research has been constrained in exploring the impact of cooperation between two levels of interaction on personalization and healthiness.
To pave the way for meal recommendation research, we introduce a new benchmark dataset called MealRec$^+$. 
Due to constraints related to user health privacy and meal scenario characteristics, the collection of data that includes both meal-course affiliation and two levels of interactions is impeded. 
Therefore, a simulation method is adopted to derive meal-course affiliation and user-meal interaction from the user's dining sessions simulated based on user-course interaction data. 
Then, two well-known nutritional standards are used to calculate the healthiness scores of meals. 
Moreover, we experiment with several baseline models, including separate and cooperative interaction learning methods. 
Our experiment demonstrates that cooperating the two levels of interaction in appropriate ways is beneficial for meal recommendations. 
Furthermore, in response to the less healthy recommendation phenomenon found in the experiment, we explore methods to enhance the healthiness of meal recommendations. 
The dataset is available on GitHub (https://github.com/WUT-IDEA/MealRecPlus).

\end{abstract}

%


\maketitle

\section{Introduction}

Health-related recommender systems profoundly impact human life, health, and well-being, especially in crucial domains such as food \cite{food_computing,cikm_liming,food_rec_1,food_framework,recipe_healthiness,category_bundle}, physical activities \cite{meal_bundle,exercise_rec_1}, healthcare \cite{healthcare_rec_1,healthcare_rec_2}, and elderly care \cite{elderly_rec_1,elderly_rec_2}. 
Beyond serving as information filters, health-related recommender systems play a pivotal role in empowering individuals to lead healthier lifestyles~\cite{health_rec_survey}. 
They serve as catalysts for promoting healthier eating habits, encouraging regular physical activity, facilitating access to appropriate healthcare resources, and fostering improved care for the elderly. 
As such, they represent a promising avenue for harnessing technology to address pressing societal challenges related to health and well-being, which aligns with the principle of AI for good \cite{ai4good_1,ai4good_2,ai4good_3}.

\begin{figure}[h]
\includegraphics[width=7.8cm]{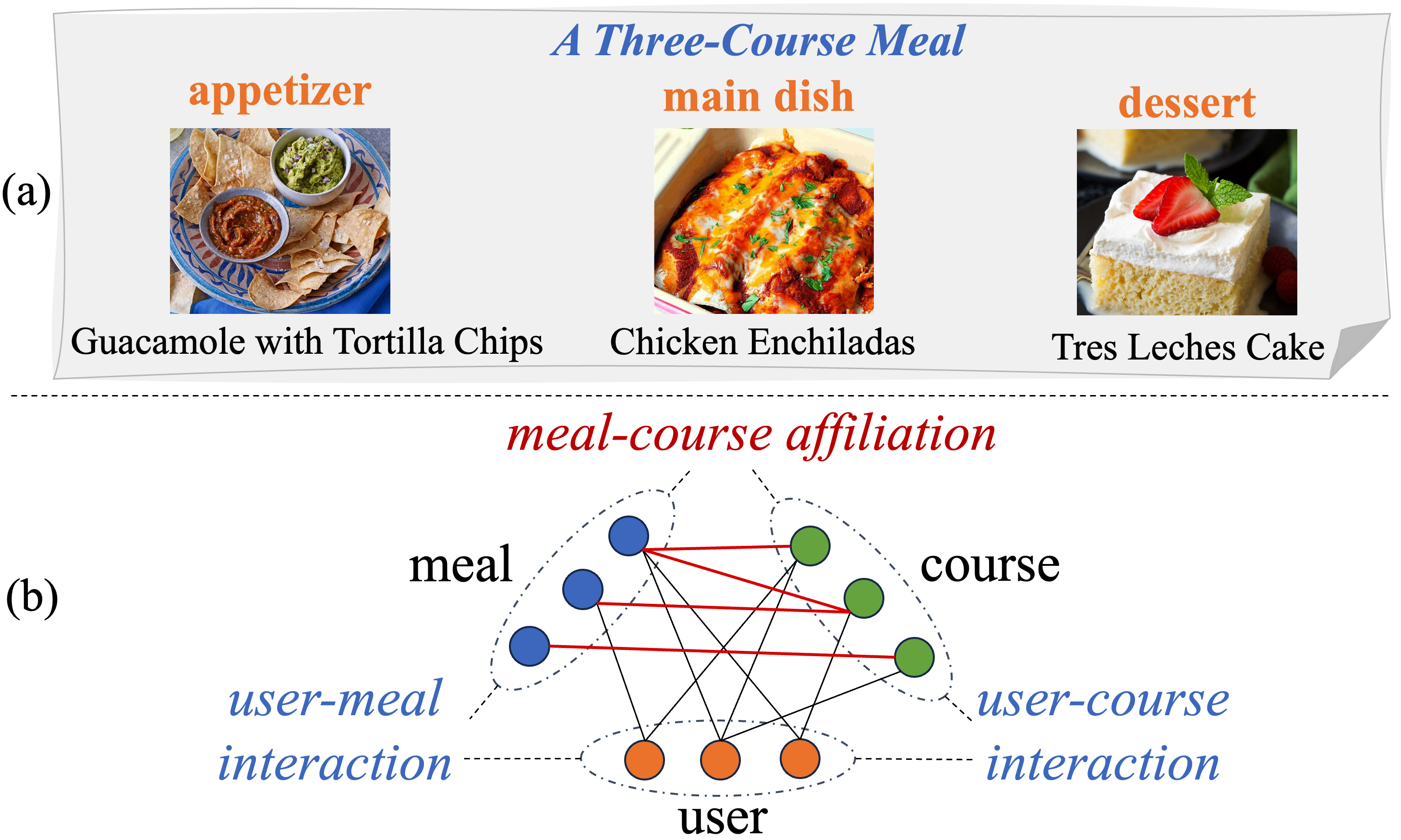}
\caption{(a) An example of three-course meals. (b) Complex relationships among users, courses, and meals.}
\label{fig-example}
\end{figure}

Eating meal is a daily activity, and meal recommender systems can guide individuals towards choices that align with their preferences,  nutritional requirements, and others. 
A meal, as a bundle of courses, usually consists of multiple courses, each from a specific category. The basic three-course meal\footnote{www.collinsdictionary.com/us/dictionary/english/three-course-meal}, as shown in Figure \ref{fig-example}(a), consists of courses from three distinct categories: appetizer, main dish, and dessert. 
As shown in Figure \ref{fig-example}(b), there are complex relationships among users, courses, and meals in the meal recommendation scenario: user-course interaction, user-meal interaction, and meal-course affiliation. 
User-course interactions contain course-level user preferences for the course features, as shown in Figure~\ref{fig-example}(a) with chicken as the main dish. 
User-meal interactions contain meal-level user preferences for the semantics expressed by the meal as a whole, as shown in Figure~\ref{fig-example}(a) is a Mexican-style meal. 
Different levels of interactions include the healthiness latent factor of user preferences.
\textbf{The meal-course affiliation can act as a bridge to connect two different levels of interaction} because they can be used to aggregate courses into meals. It corresponds to the bundle-item affiliation in bundle recommendation.

\begin{figure}[t]
\includegraphics[width=8.5cm]{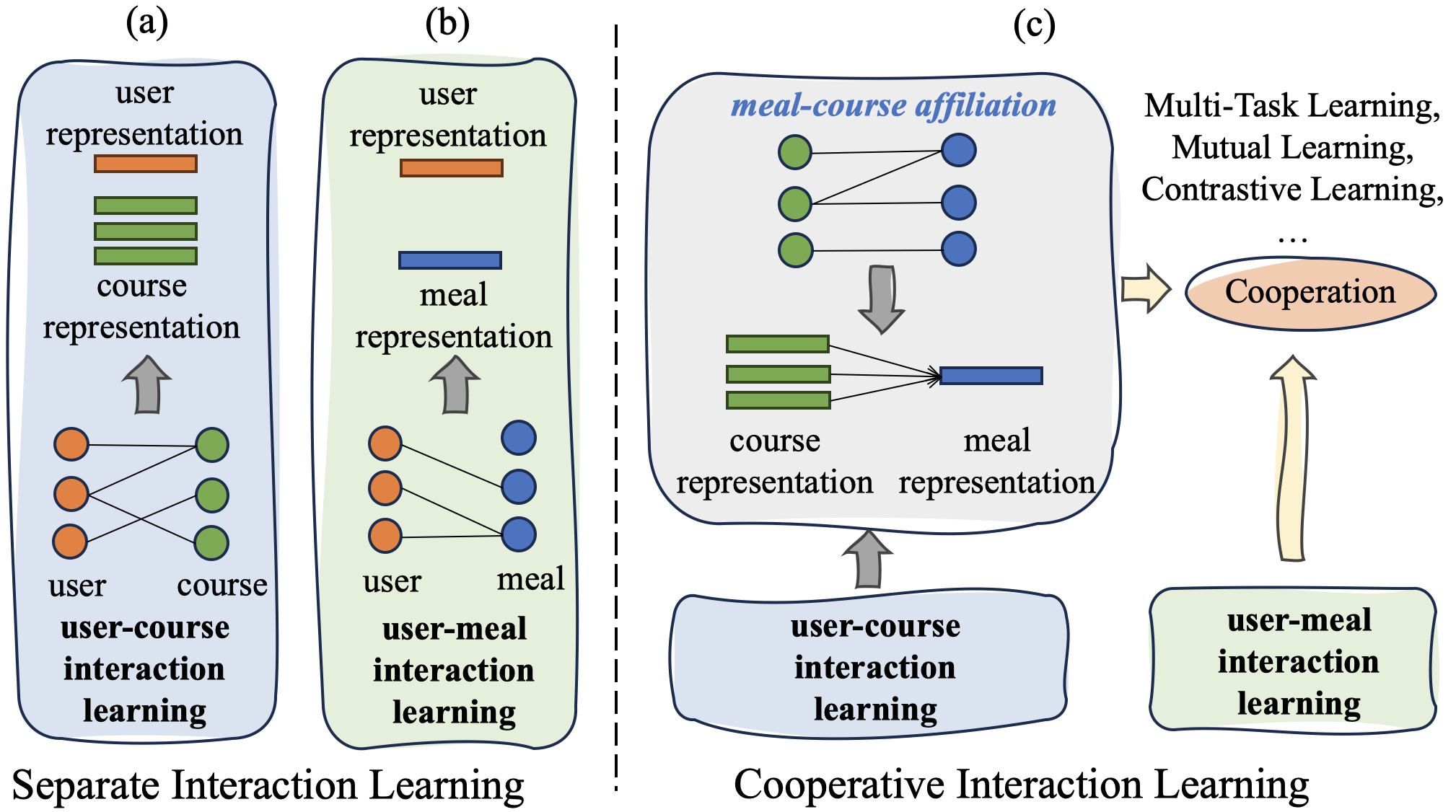}
\caption{Different interaction learning methods.}
\label{fig-combine}
\end{figure}

However, an extensive literature review demonstrates \textbf{ a lack of publicly available meal recommendation datasets with meal-course affiliation. }
This data gap leads to existing meal recommendation research being limited to separate interaction learning on different levels of interaction.  
When focusing on the user-course interaction (Figure~\ref{fig-combine}(a)), existing methods learn the user's preference for the course and then select courses that the user may like to form a meal for meal plan recommendation~\cite{meal_Elsweiler_1,meal_Elsweiler_2,meal_Yum_me}. 
When focusing on the user-meal interaction (Figure~\ref{fig-combine}(b)), existing methods recommend the meal as an ``item"~\cite{food_framework}. 
When having meal-course affiliation, cooperation between two separate interaction learnings is possible. 
As shown in Figure~\ref{fig-combine}(c), according to the meal-course affiliation, the course representations learned from the user-course interaction can be aggregated as the meal representation, so that the user preferences learned in the two levels of interactions can be trained under the same target. 
\textbf{However, due to the data gap, exploring the impact of cooperation between two levels of interaction is constrained.}

We investigate previous efforts on meal recommendation dataset construction. 
The main reason for the long-term lack of datasets containing both two levels of interactions and meal-course affiliation is that \textbf{the two main methods of collecting datasets are hindered in meal recommendation scenarios. }
1) For collecting existing real-life data, even though meal-course affiliation is public on some food-sharing websites, the user-meal interaction data is not available due to it being closely related to the user's health privacy. 
2) For collecting in an experimental environment, although eating meals is a daily activity, it cannot be repeated many times in a day, which makes collecting real meal consumption data  very time-consuming and costly. 
Similar limitations in the collection of datasets also appear in the field of conversational recommender systems. They have explored ways to simulate user dialogues to construct datasets for research~\cite{CRSs_1,CRSs_2,CRSs_3}. 

\textbf{To pave the way for meal recommendation studies, we introduce a meal recommendation benchmark dataset called MealRec$^+$.} 
Specifically, considering the category-constraints feature of meals, a mapping from course name phrases to course categories is established. 
Then, inspired by research on conversational recommender systems, meal-course affiliation and user-meal interaction are derived from user’s dining sessions simulated based on user-course interaction data from a recipe-sharing website\footnote{https://www.allrecipes.com}. 
Based on the principle of collaborative filtering that similar users may have similar behaviors, user-meal interactions are mined to expand the scale of data. 
Finally, two well-known nutritional standards from the World Health Organization and the United Kingdom Food Standards Agency are used to evaluate healthiness of meals. 
The healthiness analysis of the dataset finds that 1) users have different preferences for healthiness at course-level and meal-level, and 2) users prefer less healthy meals.   
To explore the impact of different separate and cooperative interaction learning methods on meal recommendation, we experiment with several state-of-the-art models and establish baseline performances. 
Our experiment demonstrates that cooperating the two levels of interaction in appropriate ways is beneficial for meal recommendations. 
Furthermore, in response to the less healthy recommendation phenomenon found in the experiment, we explore potential methods to improve the healthiness of recommendation results. 
In the end, we point out possible directions for building personalized and healthy meal recommender systems, with the long-term goal of transforming how users choose food conveniently and healthily. This aligns with AI for good, viz., utilizing technology’s potential to promote public good health and well-being \cite{ai4good_1, ai4good_2, ai4good_3}. 

\textbf{Our main contributions lie two-fold}: 1) We introduce a new meal recommendation benchmark dataset, MealRec$^+$, which fills a data gap by including meal-course affiliation and healthiness information. It provides conditions for researching cooperative interaction learning methods on meal recommendation, and serves as a valuable resource for AI for Good research in the food domain. 
2) Our experimental study investigates separate and cooperative interaction learning methods and potential ways for improving the healthiness in meal recommendation, which provides useful insights into diet-based public health promotion. 

\begin{table*}[th]
\small
\caption{Datasets in Meal Recommendation Research. The value in \ding{51}(\ ) represents the quantity}
\label{tab-dataset}
\begin{tabular}{c|c|c|c|c|cccc|c}
\hline 
\begin{tabular}[c]{@{}c@{}}Data Collection\\ Method\end{tabular} & Ref.                                              & Year & Source                          & \begin{tabular}[c]{@{}c@{}}Nutritional \\ Data \end{tabular} & Course & \begin{tabular}[c]{@{}c@{}} User-Course\\ Interaction\end{tabular} & \begin{tabular}[c]{@{}c@{}} Meal-Course\\ Affiliation\end{tabular} & \begin{tabular}[c]{@{}c@{}} User-Meal\\ Interaction\end{tabular} & \begin{tabular}[c]{@{}c@{}}Open\\ Source\end{tabular} \\ \hline
\multirow{5}{*}{\begin{tabular}[c]{@{}c@{}}collecting existing \\ real-life data\end{tabular}} 
& \cite{meal_recipe_similarity_2011} & 2011 & internet food sharing websites  & \ding{55}                                                               & \ding{51}(6,886)     & \ding{55}                                                                  & \ding{55}     & \ding{55}                                                                & \ding{55}                                                 \\
& \cite{meal_ingredient}               & 2012 & internet food sharing websites  & \ding{55}                                                               & \ding{51}(226,025)   & \ding{55}                                                                  & \ding{51}(unknown)   & \ding{55}                                                                & \ding{55}                                                                  \\
& \cite{meal_SousChef}                 & 2017 & official food data, nutritionist & \ding{51}                                                  & \ding{51}(1000+)     & \ding{55}                                                                  & \ding{51}(unknown)     & \ding{55}                                                                & \ding{55}                                                \\
& \cite{meal_Yum_me}                  & 2017 & internet food sharing websites  & \ding{51}                                                  & \ding{51}(75,750)    & \ding{51}(unknown)                                                              & \ding{55}     & \ding{55}                                                                & \ding{51}                                                \\
& \cite{meal_Pantry}                   & 2018 & internet food sharing websites  & \ding{51}                                                  & \ding{51}(12,930)   & \ding{55}                                                                  & \ding{55}     & \ding{55}                                                                & \ding{55}                                                \\ \hline                                                               \\ \hline 
\multirow{4}{*}{\begin{tabular}[c]{@{}c@{}}collecting in an\\ experimental \\ environment\end{tabular}}
& \cite{meal_Elsweiler_1}             & 2015 & self-built websites             & \ding{51}                                                  & \ding{51}(957)       & \ding{51}(4,549)                                                                & \ding{55}     & \ding{55}                                                                & \ding{55}                                                                  \\
& \cite{meal_DIETOS}                   & 2017 & questionnaire                   & \ding{51}                                                  & \ding{51}(unknown)   & \ding{55}                                                                  & \ding{55}     & \ding{55}                                                                & \ding{55}                                                                  \\
& \cite{meal_bundle}                   & 2019 & questionnaire                   & \ding{51}                                                  & \ding{51}(unknown)   & \ding{51}(unknown)                                                              & \ding{55}     & \ding{55}                                                                & \ding{55}                                                \\
& \cite{meal_rec_2023}                & 2023 & questionnaire            & \ding{51}                                                  & \ding{51}(40)        & \ding{51}(1,220)                                                                & \ding{55}     & \ding{55}                                                                & \ding{55}                                                                  \\ \hline
\end{tabular}
\end{table*}

\section{Related Work}
Health-related recommender systems have been receiving sustained attention from the research community\footnote{https://healthrecsys.github.io/2020/} due to their significant impact on public lives. 
As an important task among it, food recommendation helps people make healthier food decisions to reduce the risk of chronic diseases such as obesity and diabetes~\cite{food_computing}. 
Course recommendation provides a course to the user and has been widely studied~\cite{food_rec_1,food_rec_2,recipe_healthiness}. 
Meal recommendation aims to provide users with meals consisting of multiple courses, that aligns with people's natural dining habits. There are many factors that influence user choices, such as user and meal descriptions, culture, religion, tradition, price, etc., and this paper focuses on user preferences and health that health-related recommender systems are concerned about. 

\subsection{Meal Recommendation Methods}

One type of meal recommendation task is to recommend meals predefined by merchants or users.
Some existing work treats the meal as an ``item" and learns meal-level user preferences from user-meal interactions. For example, CFR~\cite{food_framework} uses MF-based collaborative filtering to learn user and meal latent vectors from the user-meal interaction matrix. 
Although recommending predefined meals are very common in real-life scenarios and are widely used online and offline, related research is limited by datasets and progresses slowly. 
Another type of meal recommendation task is meal plan recommendation, that is, selecting multiple courses to form a meal recommendation for users.  
Existing methods mainly learn course-level user preferences from user-course interactions, such as analyzing the user's rating of the course \cite{meal_bundle,meal_Yum_me}, descriptions \cite{meal_recipe_similarity_2011} and high-frequency ingredients \cite{meal_Elsweiler_1,meal_Elsweiler_2} of courses.

In short, existing methods mainly focus on learning user preferences in separate interactions. 
However, the relationship among users, courses, and meals in the meal scenario is complex. The cooperation between the two levels of interaction has yet to be fully exploited, which leaves room for improving the quality of user preference representation. 
Bundle recommendation aims to recommend a set of multiple items~\cite{bundle_first} but without considering the domain context. Its exploration on cooperation at different levels of interaction is more in-depth. Their applications of various methods, such as mutual learning~\cite{HyperMBR} and contrastive learning~\cite{CrossCBR}, are worthy of potential reference for meal recommendation. 
The research of bundle field shows that bundle-item affiliation is a necessary condition for cooperation at different levels, which corresponds to the meal-course affiliation in meal recommendation. 
Next, we will analyze the current status of data in meal recommendation.

\subsection{Meal Recommendation Datasets}
We investigate the datasets used in existing meal recommendation studies, as shown in Table~\ref{tab-dataset}. 
Among these datasets, only two contain meal-course affiliation but do not contain user interaction data. These two works~\cite{meal_ingredient,meal_SousChef} do not learn user preferences from user interactions but are based on the user’s nutritional needs or existing ingredients for meal recommendation. 
Other datasets in Table~\ref{tab-dataset} do not include meal-course affiliation, so they can only learn course-level user preferences from user-course interaction to make meal plan recommendations. 
This literature review demonstrates a lack of the dataset containing both user-course and user-meal interaction, and meal-course affiliation, constraining the research of cooperative interaction learning in meal recommendation.

To search for reasons for the long-term absence of datasets containing both interactions and meal-course affiliation, we investigate the methods of dataset collection in existing works.
As shown in the first column of Table~\ref{tab-dataset}, data collection methods are mainly divided into two categories. Some works collect existing real-world data as experimental data, and the main source is Internet food sharing websites~\cite{meal_recipe_similarity_2011,meal_ingredient,meal_Yum_me,meal_Pantry}.
Even though meal-course affiliation is publicly released on some websites, user-meal interaction can not be obtained. 
Since the user’s dietary data is closely related to the user’s health privacy, companies are very cautious about disclosing this part of the data. 
This situation also appears in the officially released data~\cite{meal_SousChef}. 
Some studies build various experimental environments and allow volunteers to participate in collecting generated user data, including self-built food-sharing websites~\cite{meal_Elsweiler_1,meal_Elsweiler_2} and questionnaires~\cite{meal_DIETOS,meal_bundle,meal_rec_2023}. This approach is resource-intensive, time-consuming and costly. 
Elsweiler et al.~\cite{meal_Elsweiler_1} built a food-sharing website to collect volunteers' meal consumption data. 
A total of 4,549 user-course interactions were collected over the four years of operation, illustrating the method has a long cycle time and is difficult to immediately assist in current research.

In general, an extensive literature review demonstrates a lack of publicly available meal recommendation datasets with meal-course affiliation. 
The main reason is that the two main data collecting methods are hindered  due to user health privacy and meal scenario characteristics. 
To create conditions for advancing representation learning in meal recommendation research, we construct a new meal recommendation dataset called MealRec$^+$. 
A simulation approach is adopted to construct MealRec$^+$, which is inspired by the method of simulating user dialogues to build datasets in conversational recommender system research \cite{CRSs_1,CRSs_2,CRSs_3}.

\section{Construction of the Dataset}
In this section, we build an implicit feedback dataset of meal recommendation. 
Dataset construction is divided into four steps: 1) Considering the category constraints characteristics of the meal, a mapping from recipe name phrases to course categories is established to lay the foundation for dataset construction. 2) Meal-course affiliations and user-meal interactions are derived from simulated user's dining sessions. 3) Next, user-meal interactions are mined based on the principle of collaborative filtering to expand the scale of data. 4) To facilitate research on health-aware meal recommendation, two widely used nutrition-based health evaluation standards are utilized to calculate the healthiness scores of meals. Finally, we analyze the constructed dataset. 

In order to find the base dataset for constructing our dataset, we search extensively in the literature and the Internet. 
Finally, we find two open-source large-scale food recommendation datasets, foodRecSys-V1\footnote{https://www.kaggle.com/elisaxxyga/foodrecsysv1} and Food180K\footnote{https://www.kaggle.com/datasets/shuyangli94/food-com-recipes-and-user-interactions}, which come from two very popular food sharing websites$^{2,}$\footnote{https://www.food.com} respectively. These two datasets have similar characteristics, such as having similar data sizes and rich course description information. The main difference is that the data density of foodRecSys-V1 is three times that of Food180K, and Food180K contains category tags, but foodRecSys-V1 does not. In our practice, data density is crucial to the construction of meal datasets, so we finally choose foodRecSys-V1 as our base dataset and use the course category tags in Food180K to build a mapping from course name to course category. foodRecSys-V1 contains 49,698 courses and 3,794,003 user-course interactions. The structure of the base dataset is shown in Table \ref{tab-org_file}. The course data contains rich descriptive information including nutrition. The interaction data includes timestamps of each user’s reviews on courses. 

\begin{table}[htb]
\caption{Description of the base data.}
  \label{tab-org_file}
  \scriptsize
  \centering
\begin{tabular}{lll}
\hline
\multicolumn{3}{c}{ course data in the base dataset} \\ \hline
Field        & Description                                                                                                      & Example                   \\ \hline
course\_id   & course identifier                                                                                                & 7994                      \\
course\_name & the name of the course                                                                                           & Coconut Poke Cake         \\
aver\_rate   & average user ratings(five-point scale)                                                                           & 4.63                      \\
image\_url   & the url of the course image                                                                                      & http://images.media...    \\
ingredients  & ingredients included in the course                                                                               & white cake mix; cream ... \\
instructions & food making process                                                                                              & Prepare and bake cake ... \\
nutrition    & nutritional content information                                                                                  & Sugars,amount:36.66 ...   \\
review\_num  & total number of the course’s reviews                                                                             & 756                       \\
reviews      & all reviews of the course                                                                                        & rating:5,text:This cake ...    \\ \hline \\ \hline

\multicolumn{3}{c}{user-course interaction in the base dataset}                                                                                                               \\ \hline
Field        & Description                                                                                                      & Example                   \\ \hline
user\_id     & user identifier                                                                                                  & 168192                    \\
course\_id   & course identifier                                                                                                & 7994                      \\
rating       & \begin{tabular}[c]{@{}l@{}}user ratings for the course\\ (five-point scale)\end{tabular}                         & 5                         \\
timestamp    & the time last modified                                                                                           & 2014-04-25T14:54:20       \\ \hline
\end{tabular}
\end{table}

\subsection{Course Category Mapping} 
First, we build a mapping table from course name phrases to categories to determine the category of course in the base dataset foodRecSys-V1. 
The course data of Food180k contains the tag field, which contains the course category tag. 11 course category tags are filtered out, which are: appetizers, main dishes, desserts, side dishes, beverages, breads, salads, snacks, soups, source, and vegetables. 
As shown in Eq.~(\ref{phrase}), each category phrase needs to meet five conditions. 
For each category, the steps to select phrases are as follows: 1) select courses based on the category label and use their course name as the corpus for that category ($origin(x)$); 2) remove stop words from the course name, such as eye-catching titles (e.g., To Die For) and version numbers (e.g., II, VI), etc ($stop\_words(x)$); 3) obtain the one-, two-, three-, and four-gram phrases of all course names; 4) count phrase frequencies, and retain the top 100 with a frequency of more than 10 of different phrases ($frequency(x)$); 5) aggregate them as the preliminary course name phrases of this category;  
6) utilize ChatGPT initially for reviewing phrases that may not distinctly represent the category ($ChatGPT\_review(x)$); 7) conduct a manual screening to remove phrases with incorrect segmentation and identify phrases that ambiguously reflect the intended category, such as phrases that appear multiple times in different categories ($manual\_screening(x)$).
\begin{equation} \label{phrase}
\begin{aligned}
\{x \in category\ phrase \mid origin(x), stop\_words(x), frequency(x), \\ ChatGPT\_review(x), manual\_screening(x)\}
\end{aligned}
\end{equation}

Based on the mapping table between course name phrases and categories, some courses in the foodRecSys-V1 dataset are assigned categories. 
Before constructing meal data, we need to clarify the composition of a meal. 
The basic meal$^{1}$ comprises three courses: an appetizer, a main dish, and a dessert.
In this paper, we refer to the composition of the three-course meal as these three categories have the most courses in the base dataset. 

\subsection{Dining Session Simulation}
As mentioned in section 2.2, the two main methods of collecting data 
do not work well in the meal scenario.
Inspired by the method of simulating user dialogues to build datasets in conversational recommender system research \cite{CRSs_1,CRSs_2,CRSs_3}, 
meal-course affiliation and user-meal interaction are derived from simulated user dining sessions based on user-course interaction. 
First, we need to clarify that the base data contains course and user-course interaction. Each course corresponds to a category. Each user-course interaction has a timestamp. 
In real life, the multiple courses consumed by a user during a dining session collectively constitute a meal, and such composition of a meal is based on the user's specific preferences and is subjectively reasonable. This user dining session can be abstracted into all course interactions of a user within a period, which can be simulated through existing user-course interactions. 

\begin{algorithm}[tbh]
\small
\SetAlgoLined
\KwIn{Base dataset with course and user-course interaction data, each course's category, user-course interaction timestamps}
\KwOut{meal-course affiliation and user-meal interactions}

$meal \gets \{\}$\;
$user\_meal\_interaction \gets \{\}$\;
\ForEach{user $u_i$}{
    $S_i \gets  simulate\_user\_dining\_session(u_i$, user-course interaction, the length of user dining session)\;
    \ForEach{user dining session s \textbf{in} $S_i$}{
    	$derived\_meal \gets derive\_meal(s$, meal composition)\;
        \If{len(derived\_meal) > $1$}{
            $probabilistically\_retain(derived\_meal)$\;
        }
        $meal.add(derived\_meal)$\;
        $user\_meal\_interaction.add$($u_i$, $derived\_meal$)\;
    }
}
$unique(meal, user\_meal\_interaction)$\;

\textbf{return} $meal$, $user\_meal\_interaction$\;

\caption{Deriving meal-course affiliation and user-meal interactions from simulated user dining sessions}
\label{algo:simulate}
\end{algorithm}

Algorithm \ref{algo:simulate} shows how user dining sessions are simulated to derive meal-course affiliation and user-meal interaction from user-course interaction. 
Note that the meal data in Algorithm \ref{algo:simulate} includes the meal-course affiliation data.
Specifically, we consider the user’s course interactions within a period of time simulated as a user dining session and all the courses in this session as candidate courses (line 4). Based on the meal's composition and the course's category, courses are selected from the candidate courses to form the meal in permutations (line 6). 
When a simulated user dining session can derive multiple meals, we randomly retain some of them to reduce similar meals (line 8). 
In addition, a user naturally interacts with the meals in his simulated dining session, and this interaction is derived in line 11. 
Upon exhaustively leveraging all simulated user meal sessions, meals with the same courses are ensured to be unique (line 14). The interaction data obtained through simulation is sparse and needs further expansion.

\subsection{Collaboration-based Interaction Mining}

The principle of collaborative filtering is that similar users are more likely to have similar behaviors. As shown in the Eq.~(\ref{similarity}), where $B(u)$ represents the user's behavior and $similarity()$ is the user similarity measure. The likelihood that two users will have similar behavior is proportional to how similar the two users are. 
\begin{equation}\label{similarity}
\begin{aligned}
P(B(u_i) = B(u_j)) \propto similarity(u_i,u_j)
\end{aligned}
\end{equation}
This foundational concept is extensively employed in recommender systems and has been empirically validated. 
The user-meal interactions derived by the simulation are fundamental interactions within the simulated environment. We mine interactions that may exist but have yet to be constructed based on fundamental interactions and the collaborative filtering principle. 
Specifically, as shown in Algorithm \ref{alg:interaction}, for each fundamental user-meal interaction ($u_i$ and $m_k$), if the user ($u_i$) and another user ($u_j$) both interacted with all the courses of the meal ($m_k$), it is considered that the two users are similar in the course interaction level (line 5). Based on the principle of collaborative filtering as shown in Eq.~(\ref{similarity}), the user ($u_j$) may also like the meal ($m_k$). 
Consequently, the interaction between the user and the meal is probabilistically established (line 6). This probability accounts for real-world exposure biases. 

\begin{algorithm}[h]
\small
\SetAlgoLined
\KwIn{Meal date, User-course interaction data}
\KwOut{Expanded user-meal interactions}

$fundamental\_interaction \gets$ user\text{-}meal\ interaction\;
$mined\_interaction \gets \{\}$\;
\ForEach{$(u_i, m_k)$ \textbf{in} fundamental\_interaction}{
    \ForEach{other user $u_j$}{
        \If{interacted\_with\_all\_courses($u_j, m_k$, user-course interaction)}{
        	$mined\_interaction.probabilistically\_add((u_j, m_k))$;
        }
    }
}
\textbf{return} $fundamental\_interaction \cup mined\_interaction$

\caption{User-meal interaction mining based on collaborative filtering}
\label{alg:interaction}
\end{algorithm}

\subsection{Healthiness Measuring}
The healthiness evaluation of meals is the basis for health-aware meal recommendation research. 
We follow the method of using nutritional standards to measure healthiness of food~\cite{recipe_healthiness,use_fsa_1,use_fsa_2,use_fsa_3,who_score} and take two internationally recognized standards for measuring in this paper: the United Kingdom Food Standards Agency (FSA) ``traffic light'' system for labeling food~\cite{fsa2013} and the World Health Organization (WHO) guidelines~\cite{who2003}.

The FSA standard focuses on 4 macro-nutrients: sugar, sodium, fat, and saturated fat. 
It assigns color indications to each macro-nutrient, such as green (healthy), amber, and red (unhealthy), based on their respective quantities. Following the approach proposed by Sacks et al.~\cite{fsa_score}, we assign numerical values to each color (green=1, amber=2, red=3). The individual scores for each macro-nutrient are then summed, resulting in an overall FSA healthiness score (referred to as FSA score) for courses, ranging from 4 (very healthy) to 12 (very unhealthy). 
Calculation follows Eq.~(\ref{FSA_formula}).
\begin{equation} \label{FSA_formula}
\begin{aligned}
    FSA\ score = \sum_{i=1}^{4}color\_value(macro\text{-}nutrient_i)
\end{aligned}
\end{equation}

The WHO defines 15 ranges for macro-nutrients to guide daily meal planning. We adopt the methodology proposed by Howard et al.~\cite{who_score}, which focuses on 7 most crucial macro-nutrients (proteins, carbohydrates, sugars, sodium, fats, saturated fats, and fibers). Points are accumulated if the macro-nutrient content falls within the specified range, resulting in the WHO healthiness score (referred to as WHO score). The calculation is based on Eq.~(\ref{WHO_formula}). The WHO score scale ranges from 0 to 7, where 0 signifies none of the WHO ranges are fulfilled (very unhealthy), and 7 indicates all ranges are met (very healthy).
\begin{equation} \label{WHO_formula}
\begin{aligned}
    WHO\ score = \sum_{i=1}^{7}within\_range(macro\text{-}nutrient_i)
\end{aligned}
\end{equation}

\begin{figure}[htb]
\includegraphics[width=7cm]{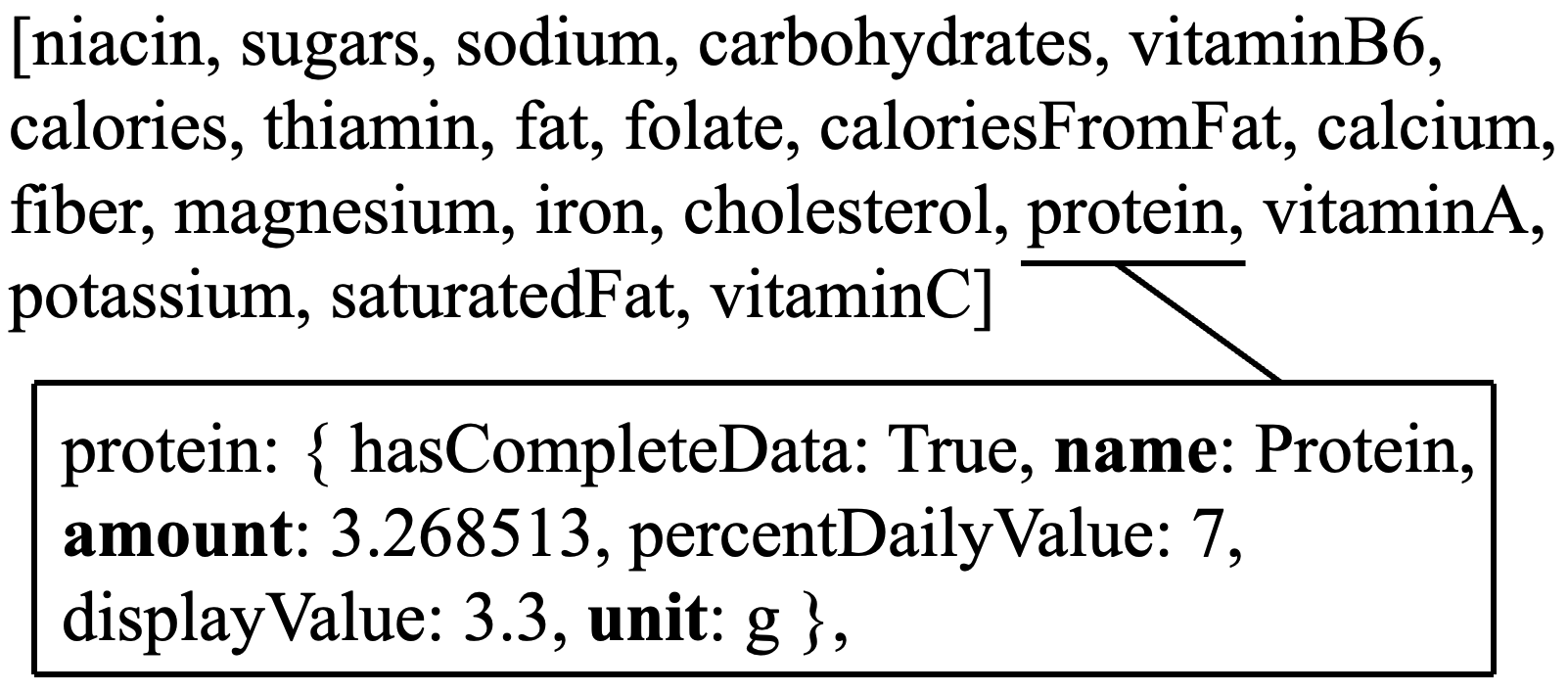}
\caption{The nutrition data format of MealRec$^+$.} 
\label{fig-nutrition}
\end{figure}

The courses in the base datatset come from the Allrecipes.com website and contain detailed nutritional data powered by the ESHA Research Database~\footnote{https://esha.com/}. The data format of nutritional data is shown in Figure~\ref{fig-nutrition}.
These nutritional data, quantified per serving, cover all macro-nutrients mentioned in two nutritional standards above. 
We calculate both the FSA score and WHO score for each course. Following other food-related studies~\cite{recipe_healthiness}, the healthiness of a meal is defined as the mean of FSA/WHO scores of its courses. 
Thus far, we have constructed the dataset, which called as MealRec$^+$.

\subsection{Dataset Analysis}

\subsubsection{Data Statistics}
\begin{figure}[h]
\includegraphics[width=8cm]{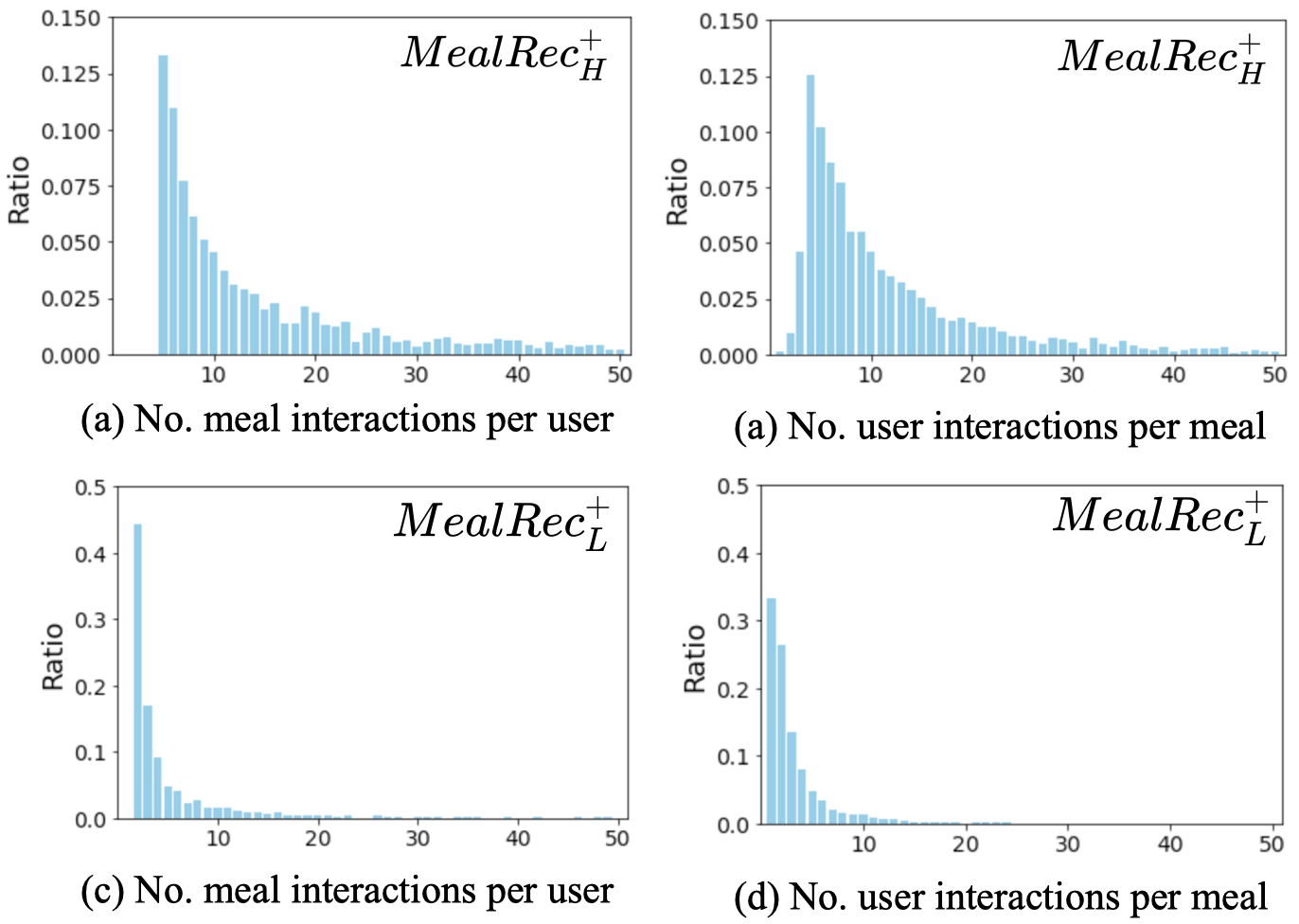}
\caption{Data distributions of datasets.}
\label{fig-distributions}
\end{figure}

Sparse filtering is performed before the constructing the dataset, and users or courses with few interactions in the base dataset are filtered. In the derived user-meal interactions, users or meals with few interactions are also filtered similarly. We construct two sub-datasets with different user-meal interaction densities, MealRec$^+_H$ with high density 0.77\% and MealRec$^+_L$ with low density 0.17\%, based on the filtering intensity of 5 and 2 interactions. 
Statistics of the two datasets are shown in Table \ref{tab:statistics}. The 7,280 courses of MealRec$^+_H$ include 2,737 appetizers, 2,552 main dishes, and 1,991 desserts. The 10,589 courses of MealRec$^+_H$ include 3,628 appetizers, 3,598 main dishes, and 3,363 desserts. 
Data distributions of MealRec$^+_H$ and MealRec$^+_L$ are shown in Figure \ref{fig-distributions}. In MealRec$^+_H$, users and meals with more than 10 interactions account for a considerable part. 
For the more sparse MealRec$^+_L$, the vast majority of users and meal interactions do not exceed 5.

\begin{table}[th]
\small
\centering
\caption{Statistics of datasets.}
\label{tab:statistics}
\begin{tabular}{l|c|c}
\hline
Dataset                & MealRec$^+_H$ & MealRec$^+_L$ \\ \hline
\# user                & 1,575    & 1,928      \\
\# meal              & 3,817    & 3,578     \\
\# course                & 7,280    & 10,589     \\
\# category            & 3       & 3        \\
\# user-meal         & 46,767   & 1,1807     \\
\# user-course           & 151,148  & 181,087    \\
\# meal-course         & 11,451   & 10,734     \\
\# user-course density   & 1.30\%  & 0.88\%   \\
\# user-meal density & 0.77\%  & 0.17\%   \\ \hline
\end{tabular}
\end{table}

Compared with other meal recommendation datasets in Table~\ref{tab-dataset}, MealRec$^+$ has both levels of interaction and meal-course affiliation, which can support the research of cooperative interaction learning methods. 
A bundle is a set of items as a whole~\cite{bundle_first}, such as a clothing outfit, a book list, a music playlist, etc. A meal is a type of bundle in the food domain. Research on bundle recommendation~\cite{BGCN,CrossCBR,HyperMBR} has emerged in recent years, and Clothing~\cite{clothing_data} and Youshu~\cite{DAM} are important benchmark datasets. 
However, they can't be applied directly to meal recommendations due to the lack of healthiness information. 
Table~\ref{tab:bundle} shows the comparison in data size between them. 
Since Youshu is one of the most important datasets of the same type, we intentionally construct the data size of MealRec$^+$ to be comparable to Youshu. 
Specifically, the length of the simulated user session affects the size of the constructed dataset, which is chosen to start with a short time and gradually extend until the data size is comparable to Youshu. 
Furthermore, Clothing and Youshu are used by existing bundle recommendation research to experiment with data-driven deep learning models. Therefore, this comparison of data size shows that MealRec$^+$ can support the applications of existing advanced recommendation technologies to meal recommendation research at the data level.

\begin{table}[t]
\centering
\small
\caption{Comparison of data size.}
\label{tab:bundle}
\begin{tabular}{c|cc|cc}
\hline
                                                                          & \multicolumn{2}{c|}{Meal Dataset} & \multicolumn{2}{c}{Bundle Dataset} \\ \cline{2-5} 
                                                                          & MealRec$^+_H$      & MealRec$^+_L$     & Clothing         & Youshu        \\ \hline
\# meal(bundle)                                                            & 3,817           & 3,578          & 1,910            & 4,771          \\
\begin{tabular}[c]{@{}c@{}}\# user-meal(bundle)\end{tabular} & 46,767          & 11,807         & 1,912            & 51,377         \\ \hline
\end{tabular}
\end{table}

\begin{table}[h]
\small
\centering
\caption{Healthiness evaluation of datasets.}
\label{tab:healthiness}
\begin{tabular}{cc|cc|cc}
\hline
                                                &           & \multicolumn{2}{c|}{MealRec$^+_H$} & \multicolumn{2}{c}{MealRec$^+_L$} \\ \cline{3-6} 
                                                &           & FSA $\downarrow$            & WHO $\uparrow$            & FSA $\downarrow$            & WHO $\uparrow$           \\ \hline
\multicolumn{2}{c|}{\begin{tabular}[c]{@{}c@{}}mean healthiness\\ score of meal\end{tabular}}                                   & 9.2758         & 2.0601         & 9.2406         & 2.0605        \\ \hline
\multicolumn{1}{c|}{\multirow{3}{*}{\begin{tabular}[c]{@{}c@{}}mean healthiness\\ score of course\end{tabular}}}      & appetizer & 8.8680         & 2.1260         & 8.9877         & 2.2138        \\
\multicolumn{1}{c|}{}                           & main dish & 9.4752         & 1.9667         & 9.4941         & 1.8471        \\
\multicolumn{1}{c|}{}                           & dessert   & 9.4841         & 2.0878         & 9.2401         & 2.1207        \\ \hline
\multicolumn{2}{c|}{\begin{tabular}[c]{@{}c@{}}mean healthiness\\ score of user interacted meal\end{tabular}}                              & 9.3652         & 2.0519         & 9.3347         & 2.0079        \\ \hline
\multicolumn{1}{c|}{\multirow{3}{*}{\begin{tabular}[c]{@{}c@{}}mean healthiness\\ score of user\\ interacted course\end{tabular}}} & appetizer & 9.0348         & 2.0553         & 9.1638         & 2.2417        \\
\multicolumn{1}{c|}{}                           & main dish & 9.4910         & 1.9863         & 9.5560         & 1.7392        \\
\multicolumn{1}{c|}{}                           & dessert   & 9.5698         & 2.1143         & 9.2843         & 2.0429        \\ \hline
\end{tabular}
\end{table}

\subsubsection{Healthiness Evaluation of Dataset} \label{health_analysis}

Mean healthiness scores of meals and courses are shown in Table~\ref{tab:healthiness}. 
$\uparrow$ indicates that the value is proportional to the health, and $\downarrow$ vice versa. we have two following observations: 

\textbf{1) User preferences for healthiness are different in user-meal interaction and user-course interaction.} The mean healthiness score of user interacted meal is different from that of user interacted course. For example, in MealRec$^+_H$, the mean FSA scores of the three categories of courses that users interacted with are 9.0348, 9.4910, and 9.5698, which is different from 0.9652 of the user interacted meal. This shows that for user preferences regarding healthiness, interaction learning of different levels of interaction captures different aspects of the user.  

\textbf{2) Users prefer less healthy meals in the MealRec$^+$ dataset.} 
Across the two datasets and two nutritional standards, it is evident that the healthiness of meals favored by users is lower than that of overall meals. For example, in MealRec$^+_L$, the mean FSA score of users, 9.3347, exceeds that of the meals, 9.2406. 
Existing models usually learn from user historical data to make recommendations. Without the guidance of healthiness in meal recommendations, the model will follow the less healthy meal preferences of some users, which will have a negative impact on the long-term health of these users. This less healthy recommendation phenomenon also emphasizes the necessity of health-aware meal recommendations.

\section{Task Definition}
Although MealRec$^+$ is also suitable for meal plan recommendation, we focus on recommending predefined meals because their research is relatively lagging but they have great application value in reality. 
The impact of other factors (such as user or meal description, culture, religion, tradition, price, etc.) and meal plan recommendation task are left as future work.
The task definition of meal recommendation in this paper can be formulated as follows:

\textbf{Input:} Users, meals, courses, categories, healthiness of courses and meals, user-meal interaction data, user-course interaction data, meal-course affiliation data, course-category corresponding data. 

\textbf{Output:} A personalized and health-aware scoring function that maps a meal $m$ to a real value for each user: $f_u:m\to \mathbb{R}$.

\section{Baselines}
This section introduces potential separate and cooperative interaction learning methods. 
Existing meal recommendation methods mainly use separate interaction learning, so we choose one of them, CFR~\cite{food_framework}, as our baseline. 
Meal recommendation is a food-domain subtask of bundle recommendation which aims to present a user with a set of multiple items~\cite{bundle_first}. 
Compared with meal recommendation research, bundle recommendation research~\cite{BGCN,MIDGN,CrossCBR,DAM,tree_bundle,HyperMBR} has sophisticated and public  benchmark datasets and the research is more in-depth. In this paper, we introduce several state-of-the-art models in bundle recommendation as baseline models to explore the effectiveness of different interaction learning methods.

\subsection{Separate Interaction Learning Method}
\subsubsection{Learning on Two Levels of Interaction}
\textbf{BGCN}~\cite{BGCN} utilizes separate interaction learning based on graph neural network (GNN) to learn user preferences on two levels of interaction, respectively. 
For user-course interaction, BGCN learns course-level user preferences and course representations and then aggregates the course representation into a meal representation according to meal-course affiliation. 
For user-meal interaction, BGCN directly uses GNN to learn meal-level user preferences and meal representations.
BGCN calculates the inner product of user and meal representations at two levels, respectively, and naively adds them together as the final prediction result. 
However, there is no cooperation between the two levels interaction learning. 

\subsubsection{Learning on One Level of Interaction}
In order to verify the effectiveness of separate interaction learning on the two levels of interaction, we design two additional ablation models of BGCN as baseline models. \textbf{CourseLevel} conducts separate interaction learning to learn course-level user preferences from user-course interaction. \textbf{MealLevel} conducts separate interaction learning to learn meal-level user preferences from user-meal interaction. 
Moreover, \textbf{CFR}~\cite{food_framework}, a general food recommendation framework, is chosen as our baseline. It uses MF-based collaborative filtering to learn user and meal latent vectors from user-meal interaction. 

\subsection{Cooperative Interaction Learning Methods}

\subsubsection{Multi-Task Learning}
\textbf{DAM}~\cite{DAM} is an attention-based multi-task model. DAM shares user embeddings to predict both user-meal interactions and user-course interactions, and shares course embeddings to learn meal representations and to predict user-course interactions. During model learning, DAM tightly couples the models for user-meal and user-course interactions in a multi-task manner, allowing the benefits of one task (user-course modeling) to be transferred to another task (user-meal modeling).

\subsubsection{Mutual Learning}

\textbf{HyperMBR}~\cite{HyperMBR} is a hyperbolic mutual learning model. It follows BGCN's operation of learning course-level and meal-level user preferences from two levels of interaction. However, HyperMBR argues that different levels have different learning trends. Meal level tends to recommend the meals that similar users have interacted with to the target user. Course level focuses more on the meal-course similarity.
In order to allow the interaction learning of two different levels to reinforce each other, HyperMBR enables two levels to continuously extract complementary knowledge contained in soft labels through mutual learning. 

\subsubsection{Contrastive Learning}

\textbf{CrossCBR}~\cite{CrossCBR} is the state-of-the-art bundle recommendation model. It also follows BGCN to learn user preferences in two levels. 
 However, CrossCBR believes that BGCN loosely combines the predictions of two separate levels, while the crucial cooperative association between the two levels’ representations is overlooked. CrossCBR proposes to model the cooperative association between the two different levels through contrastive learning. By encouraging the alignment of the two separately learned interactions, each level can distill complementary information from the other level, achieving mutual enhancement. 

\section{Experiment on Personalization and Healthiness}

\subsection{Experiment Setup}

\subsubsection{Evaluation Protocol}
All-unrated-item\cite{evaluation} as a widely used evaluation protocol
is adopted in this paper: for each user, we retain all candidate meals with whom she does not interact within the training set. 
For both sub-datasets, we divide training, validation,
and test set in a ratio of 8:1:1. 
Top-K meal recommendation is regarded as the specific task of this paper: recommend a ranked list of K meals from the candidate meals to a user. 

\subsubsection{Metrics on personalization}
Widely used Recall@K and NDCG@K \cite{BGCN,DAM,CrossCBR} are employed to evaluate the top-K recommendation performance of personalization. Recall measures the ratio of test meals within the top-K list, and NDCG accounts for the position of the hits by assigning higher scores to those at top ranks. We follow the previous work~\cite{BGCN,CrossCBR} and take K=20,40,80. 

\subsubsection{Metrics on Healthiness} 
\textbf{1) Mean Healthiness Score}: 
Following the approach of Trattner et al.~\cite{recipe_healthiness}, the mean FSA and WHO scores (denoted as $Score_{FSA}$/$Score_{WHO}$) of the top-K recommended meals are used to evaluate the absolute healthiness of the recommendation results. 
$\triangle Score_{FSA}$ and $\triangle Score_{WHO}$ are the healthiness score difference between the top-K recommended meals and the meal the user has interacted with.
They are used to evaluate the relative healthiness of the recommended meals. As an example, the calculation formula of $\triangle Score_{FSA}$ is shown in Eq.~(\ref{FSA}), where $Score_{FSA}^{history}$ is the mean FSA score of history data. 
A negative $\triangle Score_{FSA}$ and a positive $\triangle Score_{WHO}$ indicate that the recommended list is healthier than the average of the user's history. 
\begin{equation} \label{FSA}
\begin{split}
 \triangle Score_{FSA} = Score_{FSA} - Score_{FSA}^{history}
\end{split}
\end{equation}

\textbf{2) Ranking Exposure}: 
Highly ranked meals in a list are more likely to attract user clicks, making meal ranking a crucial factor. 
Widely used metrics ranking exposure~\cite{ranking_exposure1,ranking_exposure2} is adopted to evaluate the performance of meal groups with different healthiness in ranking. 
Sorting meals based on their healthiness, we categorize the top half into the relatively healthy meal group $\mathcal{G}h$ and the bottom half into the relatively less healthy meal group $\mathcal{G}{lh}$. The ranking exposure for $\mathcal{G}_h$ under a ranking order $r$ is defined as:
\begin{equation} \label{rank_metric}
\begin{aligned}
    Expo(\mathcal{G}_h|r) = \sum_{m\in\mathcal{G}_h} utility(m|r),
\end{aligned}
\end{equation}
where $utility(\cdot)$ denotes the utility function, which is usually a monotonically decreasing function with respect to the rank of an meal. In this paper, we set $utility(m|r) = [rank(m|r)^{-1}]$~\cite{ranking_exposure2}. The ranking exposure gap between $\mathcal{G}_h$ and $\mathcal{G}_{lh}$ is defined as:
\begin{equation} \label{rank_gap}
\begin{aligned}
    ExpoGap(\mathcal{G}_h,\mathcal{G}_{lh}|r)=\frac{Expo(\mathcal{G}_h|r)-Expo(\mathcal{G}_{lh}|r)}{Expo(\mathcal{G}_h|r)+Expo(\mathcal{G}_{lh}|r)}
\end{aligned}
\end{equation}

We abbreviate ExpoGap($\mathcal{G}_h,\mathcal{G}_{lh}|r$) as ExpoGap. 
Given that the two groups share the same size, ExpoGap = 0 implies no disparity in ranking exposure between two groups. 
A positive (negative) ExpoGap means higher ranking exposure for relatively healthy (less healthy) meals, with the value indicating the extent of the gap. 

\subsubsection{Hyper-parameters Setting}
We reproduce CFR based on their reported framework design. 
Other baseline models are implemented using their open-source codes. 
The embedding size is set to 64, commonly used in models~\cite{BGCN,CrossCBR}. The mini-batch size is fixed at 1024. 
The propagation layer in the GNN-based model is stacked 2 layers. The training of all models is not limited to epochs but has a patience of 10 epochs. The epoch result with the best Recall@20 on the test set is reported as the final result of each model.
All experiments are performed on an Nvidia TITAN Xp GPU.  

\begin{table*}[t]
\small
\caption{Baseline model performance in personalization of recommendation.}
\label{tab:res-personalization}
\begin{tabular}{c|c|cccccc|cccccc}
\hline
\multicolumn{2}{c|}{\multirow{3}{*}{Model}}                                                                  & \multicolumn{6}{c|}{MealRec$^+_H$}                                                                         & \multicolumn{6}{c}{MealRec$^+_L$}                                                                         \\ \cline{3-14} 
\multicolumn{1}{c}{} & \multicolumn{1}{c|}{}                                                                                      & \multicolumn{2}{c|}{K=20}            & \multicolumn{2}{c|}{K=40}            & \multicolumn{2}{c|}{K=80} & \multicolumn{2}{c|}{K=20}            & \multicolumn{2}{c|}{K=40}            & \multicolumn{2}{c}{K=80} \\ \cline{3-14} 
\multicolumn{1}{c}{} & \multicolumn{1}{c|}{}                                                                                      & Recall$\uparrow$ & \multicolumn{1}{c|}{NDCG$\uparrow$}   & Recall$\uparrow$ & \multicolumn{1}{c|}{NDCG$\uparrow$}   & Recall$\uparrow$      & NDCG$\uparrow$        & Recall$\uparrow$ & \multicolumn{1}{c|}{NDCG$\uparrow$}   & Recall$\uparrow$ & \multicolumn{1}{c|}{NDCG$\uparrow$}   & Recall$\uparrow$      & NDCG$\uparrow$       \\ \hline
\multicolumn{1}{c|}{\multirow{4}{*}{\begin{tabular}[c]{@{}c@{}}Separate\\ Interaction\\ Learning\end{tabular}}} & CFR~\cite{food_framework}      & 0.1352 & \multicolumn{1}{c|}{0.0777} & 0.1897 & \multicolumn{1}{c|}{0.0915} & 0.2505      & 0.1057      & 0.0330 & \multicolumn{1}{c|}{0.0145} & 0.0509 & \multicolumn{1}{c|}{0.0183} & 0.0742      & 0.0229     \\                           
\multicolumn{1}{c|}{} & CourseLevel      & 0.2181 & \multicolumn{1}{c|}{0.1345} & 0.2792 & \multicolumn{1}{c|}{0.1500} & 0.3684      & 0.1706      & 0.1181 & \multicolumn{1}{c|}{0.0668} & 0.1565 & \multicolumn{1}{c|}{0.0754} & 0.2104      & 0.0858     \\
\multicolumn{1}{c|}{} & MealLevel      & 0.1450 & \multicolumn{1}{c|}{0.0812} & 0.2040 & \multicolumn{1}{c|}{0.0964} & 0.2891      & 0.1150      & 0.0514 & \multicolumn{1}{c|}{0.0194} & 0.0776 & \multicolumn{1}{c|}{0.0250} & 0.1187      & 0.0326     \\ 
\multicolumn{1}{c|}{}       & BGCN~\cite{BGCN}     & 0.2083 & \multicolumn{1}{c|}{0.1263} & 0.2787 & \multicolumn{1}{c|}{0.1440} & 0.3750      & 0.1653      & 0.0882 & \multicolumn{1}{c|}{0.0438} & 0.1130 & \multicolumn{1}{c|}{0.0496} & 0.1809      & 0.0624     \\ \hline 
\multicolumn{1}{c|}{\multirow{3}{*}{\begin{tabular}[c]{@{}c@{}}Cooperative\\ Interaction\\ Learning\end{tabular}}}   
& DAM~\cite{DAM} & 0.2043 & \multicolumn{1}{c|}{0.1266} & 0.2773 & \multicolumn{1}{c|}{0.1450} & 0.3640      & 0.1648      & 0.0986 & \multicolumn{1}{c|}{0.0498} & 0.1366 & \multicolumn{1}{c|}{0.0581} & 0.2018      & 0.0706     \\                                                
\multicolumn{1}{c|}{}                      & HyperMBR~\cite{HyperMBR} & 0.3144 & \multicolumn{1}{c|}{0.2153} & 0.4240 & \multicolumn{1}{c|}{0.2436} & 0.5359      & 0.2702      & 0.1497 & \multicolumn{1}{c|}{0.0786} & 0.1998 & \multicolumn{1}{c|}{0.0894} & 0.2757      & 0.1034     \\ 
\multicolumn{1}{c|}{}                                                                            & CrossCBR~\cite{CrossCBR} & 0.3753 & \multicolumn{1}{c|}{0.2588} & 0.4825 & \multicolumn{1}{c|}{0.2868} & 0.6053      & 0.3156      & 0.1920 & \multicolumn{1}{c|}{0.1018} & 0.2349 & \multicolumn{1}{c|}{0.1116} & 0.3055      & 0.1248     \\ \hline
\end{tabular}
\end{table*}

\begin{table*}[h]
\small
\caption{Baseline model performance in healthiness of recommendation(K=20, FSA nutritional standard).}
\label{tab:res-healthiness}
\begin{tabular}{c|c|cc|cc}
\hline
\multicolumn{2}{c|}{\multirow{3}{*}{Model}}                                                                 & \multicolumn{2}{c|}{MealRec$^+_H$}          & \multicolumn{2}{c}{MealRec$^+_L$}           \\ \cline{3-6} 
\multicolumn{1}{c}{} & \multicolumn{1}{c|}{}                                                                                       & \multicolumn{1}{c|}{Mean Healthiness Score} & Ranking Exposure & \multicolumn{1}{c|}{Mean Healthiness Score} & Ranking Exposure \\ \cline{3-6}
\multicolumn{1}{c}{} & \multicolumn{1}{c|}{}                                                                                       & \multicolumn{1}{c|}{$Score_{FSA}$($\triangle$)$\downarrow$}         & ExpoGap$\uparrow$         & \multicolumn{1}{c|}{$Score_{FSA}$($\triangle$)$\downarrow$}         & ExpoGap$\uparrow$         \\ \hline 
\multicolumn{1}{c|}{\multirow{4}{*}{\begin{tabular}[c]{@{}c@{}}Separate\\ Interaction\\ Learning\end{tabular}}} & CFR~\cite{food_framework}     & \multicolumn{1}{c|}{9.4611(0.0959)}         & -0.1861         & \multicolumn{1}{c|}{9.3720(0.0373)}         & -0.2086         \\                                                                   
\multicolumn{1}{c|}{} & CourseLevel     & \multicolumn{1}{c|}{9.4384(0.0732)}         & -0.1761         & \multicolumn{1}{c|}{9.4037(0.0690)}         & -0.2136         \\
\multicolumn{1}{c|}{} & MealLevel      & \multicolumn{1}{c|}{9.4827(0.1175)}         & -0.2311         & \multicolumn{1}{c|}{9.3881(0.0534)}         & -0.2348         \\
\multicolumn{1}{c|}{}       & BGCN~\cite{BGCN}     & \multicolumn{1}{c|}{9.4402(0.0750)}         & -0.1718         & \multicolumn{1}{c|}{9.4115(0.0768)}         & -0.2686         \\ \hline
\multicolumn{1}{c|}{\multirow{2}{*}{\begin{tabular}[c]{@{}c@{}}Cooperative\\ Interaction\\ Learning\end{tabular}}} 
& DAM~\cite{DAM} & \multicolumn{1}{c|}{9.4342(0.0690)}         & -0.1614         & \multicolumn{1}{c|}{9.3656(0.0309)}         & -0.1460         \\
\multicolumn{1}{c|}{}                                & HyperMBR~\cite{HyperMBR} & \multicolumn{1}{c|}{9.5458(0.1806)}         & -0.2638         & \multicolumn{1}{c|}{9.4138(0.0791)}         & -0.3010         \\
\multicolumn{1}{c|}{}                                                                            & CrossCBR~\cite{CrossCBR} & \multicolumn{1}{c|}{9.3229(-0.0423)}        & -0.0737         & \multicolumn{1}{c|}{9.3475(0.0128)}         & -0.1346         \\ \hline
\end{tabular}
\end{table*}

\begin{table*}[th]
\small
\caption{The effectiveness of hard filtering and re-weighting methods in improving healthiness of recommendations.}
\label{tab:res-filter}
\begin{tabular}{c|cccc|cccc}
\hline
\multirow{3}{*}{Model} & \multicolumn{4}{c|}{MealRec$^+_H$}                                  & \multicolumn{4}{c}{MealRec$^+_L$}                                   \\ \cline{2-9} 
                       & \multicolumn{4}{c|}{K=20}                                        & \multicolumn{4}{c}{K=20}                                         \\ \cline{2-9} 
                       & Recall$\uparrow$ & \multicolumn{1}{c|}{NDCG$\uparrow$}   & $Score_{FSA}$($\triangle$)$\downarrow$  & ExpoGap$\uparrow$ & Recall$\uparrow$ & \multicolumn{1}{c|}{NDCG$\uparrow$}   & $Score_{FSA}$($\triangle$)$\downarrow$  & ExpoGap$\uparrow$ \\ \hline
backbone model              & 0.3753 & \multicolumn{1}{c|}{0.2588} & 9.3229(-0.0423) & -0.0737 & 0.1920 & \multicolumn{1}{c|}{0.1018} & 9.3475(0.0128)  & -0.1346 \\ \hline
Hard Filtering            & 0.0682 & \multicolumn{1}{c|}{0.0582} & 7.5916(-1.7736) & 1.0000       & 0.0277 & \multicolumn{1}{c|}{0.0126} & 7.6275(-1.7072) & 1.0000       \\
Re-Weighting            & 0.2817 & \multicolumn{1}{c|}{0.1702} & 8.3673(-0.9979) & 0.7931  & 0.1544 & \multicolumn{1}{c|}{0.0731} & 8.3925(-0.9422) & 0.7273  \\ \hline
\end{tabular}
\end{table*}

\subsection{Baseline Performance} \label{result}
Baseline model performance in personalization and healthiness are shown in Table~\ref{tab:res-personalization} and Table~\ref{tab:res-healthiness}, respectively. The healthiness performance under different Top-K recommendation lists and different health standards is similar. This paper reports the results based on the FSA standard and set K=20. 
The arrow $\uparrow$ means that the larger the value, the better the performance, while $\downarrow$ means the opposite. $\triangle$ in Table~\ref{tab:res-healthiness} represents the difference in healthiness scores compared to historical data. 
We have the following observations:
 
\textbf{1) Separate interaction learning of different level interaction performs differently.} As shown in Table~\ref{tab:res-personalization}, the Recall@20 of CFR and MealLevel on MealRec$^+_H$ are 0.1352 and 0.1450, which are much lower than the 0.2181 of CourseLevel. 
We think it is mainly due to that the user-meal interaction is more sparse than the user-course interaction(see Table~\ref{tab:statistics}). 
The personalization performance of BGCN is not as good as that of CourseLevel, which shows that simply summing the prediction results without the cooperation of two levels cannot improve the effect. On the contrary, the poor performance of MealLevel drags down CourseLevel. 

\textbf{2) Cooperating two different interaction learning in appropriate ways can effectively improve the recommendation accuracy.} In Table~\ref{tab:res-personalization}, HyperMBR and CrossCBR apply cooperative interaction learning to learn user preferences from two levels of interactions and achieve 0.3144 and 0.3753 on MealRec$^+_H$. They outperform the baseline models with separate interaction learning in Recall and NDCG on both datasets. This shows that the learning of user-course interaction and user-meal interaction are not isolated, but should cooperate, complement, and enhance each other. 
DAM is not better than CourseLevel and BGCN. We think the main reason is that the representation learning ability of attention networks is not as good as that of GNN, not that the cooperation is not effective, because DAM is improved compared to CFR. 

\textbf{3) Different cooperative interaction learning methods have different impacts on the healthiness of recommendations.} As shown in Table~\ref{tab:res-healthiness}, compared with BGCN, HyperMBR uses mutual learning to cooperate interaction and reduces the healthiness of recommendations, with $Score_{FSA}$ reaching 9.5458 on MealRec$^+_H$. With contrastive learning, CrossCBR performs best in learning users' healthiness preferences, and the absolute value difference from $Score_{FSA}$ in user historical data is only 0.0423.

\textbf{4)Without health-oriented guidance, baseline models follow the user's less healthy preferences and make less healthy recommendations.} 
According to the analysis of user historical data in Section	~\ref{health_analysis}, users prefer relatively less healthy meals.
Baseline models learn the user's preferences from the user's course interaction and meal interaction. This preference representation includes health-related latent factors. Baseline models focus on learning user preferences without guiding the healthiness of recommendations. 
Subsequently, These models cater to users’ less healthy preferences and recommend less healthy meals. 
As shown in Table~\ref{tab:res-healthiness}, the FSA scores of most models are almost consistent with or worse with user historical data ($\triangle Score_{FSA} \ge 0$, the smaller the FSA score, the healthier). The ranking exposure values of all models are negative, indicating that less healthy meals have an advantage in recommended rankings. 

\section{Improve Healthiness of Recommendation} \label{sec:improve_health}
Baseline models do not provide health-oriented guidance, making recommendations following the user's less healthy preferences. 
In this paper, we try to use two post-filtering methods to improve the healthiness of recommendations: 1) \textbf{Hard Filtering}(HF), which directly filters out meals that do not meet the healthiness score before recommendation, 2) \textbf{Re-weighting}(RW)~\cite{recipe_healthiness}, which applies a scoring function which re-weights the scores of a meal for a particular user based on the FSA score of the meal. The re-weighting function is shown in Eq. (\ref{RW}), where $r_{u,m}$ is the original output result of the model, and $fsa_m$ is the FSA score of the meal. 
\begin{equation} \label{RW}
\begin{split}
r_{u,m,fsa} = r_{u,m}\cdot(13-fsa_m)
\end{split}
\end{equation}

CrossCBR is chosen as the backbone model to test the effects of the two model-agnostic methods because it achieves the best personalization performance before. Filter strength for HF is set to 8 (FSA score). 

The experimental results are shown in Table~\ref{tab:res-filter} and our observations are as follows: \textbf{1) The hard filtering method achieves the best performance on healthiness but poor on personalization}. HF filters out all meals with an FSA score above 8, and $Score_{FSA}$ is reduced to 7.5916 in MealRec$^+_H$. However, excessive intervention in the outcome results in poor personalized performance. \textbf{2) The re-weighting method improves the healthiness of recommendations while retaining a certain accuracy of recommendations}. 
RW improves the ExpoGap of the backbone model on MealRec$^+_L$ from -0.1346 to 0.7273. 
At the same time, RW's Recall@20 in MealRec$^+_L$ is 0.1544, much higher than HF's 0.0277.

\section{Conclusion and Future Directions}
In this paper, we have introduced MealRec$^+$, a meal recommendation dataset including meal-course affiliation. 
MealRec$^+$ fills the data gap and provides conditions for exploring advanced user preference representation technologies in meal recommendation research. 
We experiment with different separate and cooperative interaction learning methods and potential ways of improving healthiness, and establish baseline performances in terms of personalization and healthiness. 
This paper provides new resource and useful insights for AI for good in food domain. 
In the future, 
an important direction is to propose user interaction learning methods more suitable for meal recommendation contexts~\cite{NICF}, such as category constraint characteristics of meals~\cite{SIGIR-AP,HBPR,category_bundle} and other factors like user and meal descriptions~\cite{food_image1,food_image2}, ingredients~\cite{ingredients}, religion~\cite{religion}, availability~\cite{meal_Pantry,availability_1}, etc. It is interesting to experiment with other frameworks and recommendation methods~\cite{newRef2,newRef3,newRef4}. 
Another direction worthy of research is how to trade off the personalization and healthiness, 
which is very essential for users to accept healthy meals and develop healthy eating habits. 
Potential ways are to guide the training of the model in a health-oriented way~\cite{HBPR,trade_off1}, or to treat less healthy recommendations of the model as a bias and improve the health of the recommendations through debias methods~\cite{use_fsa_3,debias1,debias2,debias3}. 

\begin{acks}
This work is partially supported by NSFC of China (No.62276196) and NSERC of Canada. Thanks to all reviewers for their comments.
\end{acks}

\bibliographystyle{ACM-Reference-Format}
\balance
\bibliography{references}

\end{sloppypar}
\end{document}